\documentclass[aps,prb,twocolumn,superscriptaddress,showpacs,floatfix]{revtex4-1}
\usepackage{graphicx,amssymb}
\usepackage{color,ulem} 

\renewcommand{\Re}{\mathop \mathrm{Re}}
\renewcommand{\Im}{\mathop \mathrm{Im}}

\begin{document}

\title{Melting of Wigner crystal  in high-mobility $n$-GaAs/AlGaAs
heterostructures at filling factors $0.18 > \nu > 0.125$: Acoustic
studies.}

\author{I.~L.~Drichko}
\author{I.~Yu.~Smirnov}
\affiliation{A.~F.~Ioffe Physico-Technical Institute of Russian
Academy of Sciences, 194021 St. Petersburg, Russia}
\author{A.~V.~Suslov}
\affiliation{National High Magnetic Field Laboratory, Tallahassee, FL 32310, USA}
\author{Y.~M.~Galperin}
\affiliation{Department of Physics, University of Oslo, 0316 Oslo, Norway}
\affiliation{A.~F.~Ioffe Physico-Technical Institute of Russian
Academy of Sciences, 194021 St. Petersburg, Russia}
\author{L.~N.~Pfeiffer}
\author{K.~W.~West}
\affiliation{Department of Electrical Engineering, Princeton University, Princeton, NJ 08544, USA}

\begin{abstract}
Using acoustic methods the complex  high-frequency conductance of
high-mobility $n$-GaAs/AlGaAs heterostructures was determined in
magnetic fields  12$\div$18~T. Based on the observed frequency and
temperature dependences we conclude that in the investigated
magnetic field range and at sufficiently low temperatures, $T \lesssim
200$~mK, the electron system forms a Wigner crystal deformed due
to  pinning by disorder. At some temperature, which depends on the
electron filling factor, the temperature dependences of both
components of the complex conductance get substantially changed. We
have ascribed this rapid change of the conduction mechanism to
melting of the Wigner crystal and study the dependence of the so-defined
melting temperature on the electron filling factor.

\end{abstract}


\date{\today}

\maketitle

\section{Introduction} \label{introduction}

%
%
%

Transport properties of a  two-dimensional electron system (2DES) in
high magnetic fields ($B$) are governed by an interplay between
electron-electron interaction and their interaction with impurities.
Both interactions depend on the typical size of the electron wave
function, which is parameterized by the magnetic length,
$l_B=\sqrt{c\hbar/eB}$. At high $B$, $l_B \to 0$ and electrons act
as classical point particles. Without disorder, such particles tend
to form a triangular lattice -  a Wigner crystal (WC) - stabilized by
electron repulsion.~\cite{Wigner1934} The wave function overlap
decreases with increase of $B$. Its role is quantitatively
characterized by the ratio between $l_B$ and the lattice constant,
$a$, of the WC. The ratio $l_B/a$ is related to the Landau filling
factor, $\nu$, as $\nu =nh/eB =(4\pi/\sqrt{3}) (l_B/a)^2$. At
sufficiently high $\nu$ the WC ground state is predicted to undergo
a transition to the fractional quantum Hall effect (FQHE) state, see,
e.g., Ref.~\onlinecite{Shayegan2005} for a review.

Since 2DESs at high magnetic  fields are insulators it is concluded
that WC is pinned by disorder. The disorder leads to texturing of
the electron system into domains, typical size $L$  of which (the
so-called Larkin-Ovchinnikov length) can be estimated comparing the
cost in shear elastic energy and the gain due to
disorder.~\cite{Larkin1979} This conclusion is supported  by
observation of well-defined resonances in the microwave absorption
spectrum.~\cite{Jiang1991,Shayegan1997} In the pinning mode, parts
of WC oscillate within the disorder-induced  potential, which
defines the so-called pinning frequency, $\omega_p$.

These oscillations get mixed with the cyclotron motion in the magnetic field resulting in absorption peaks at some  frequencies, $f_{pk}$.~\cite{Fukuyama1978,Normand1992,Millis1994} In the classical, high-$B$ limit, where $l_B$ is much smaller than any feature of the disorder, and also small enough that the wave function overlap of neighboring electrons can be neglected,  $ f_{pk} \propto B^{-1}$.

The perfection of the WC order in 2DES has been addressed previously using time-resolved photoluminescence~\cite{Kukushkin1994,*Kukushkin1996}  provided
evidence for triangular crystalline ordering in the high-$B$ regime. In double quantum wells, evidence for ordering came from commensurability effects.~\cite{Manoharan1996} In the context of the model described in Ref.~\onlinecite{Millis1994}, the domain size has been estimated
previously from early microwave,~\cite{Glattli1990,*Williams1991} surface acoustic
wave,~\cite{Paalanen1992} and nonlinear I-V data.~\cite{Glattli1990,*Williams1991,Li1991}

Previously~\cite{Drichko2015}  we have studied  dependences of
complex conductance, $\sigma ^{AC} (\omega) \equiv \sigma_1 (\omega)
- i \sigma_2 (\omega)$, on frequency, temperature, and magnetic
field in the vicinity of the filling factor $1/5$, namely for $0.19
< \nu < 0.21$. The complex conductance was extracted from
simultaneous measurements of magnetic field dependences of
attenuation and variation of velocity of surface acoustic waves
(SAW) propagating in the vicinity of the sample surface. The results
were interpreted as evidence of formation of a pinned Wigner
crystal (WC). This conclusion was based on an observed maximum in
the frequency dependence of $\sigma_1$ at $f\equiv \omega/2\pi \sim
100$~MHz coinciding with a change of the sign of $\sigma_2
(\omega)$. This results allowed us to estimate the domain size in
the pinned WC.

In this paper we study the dependences  of complex conductance on the
frequency, the temperature and the SAW electric field intensity in the
same structure, but in higher magnetic field $12<B< 18$~T corresponding to $0.18
> \nu > 0.125$, respectively. The
measurements are made for temperatures $T=(40\div340)$~mK and SAW
frequencies $f=(30\div300)$~MHz.

The paper is organized as follows.  In Sec.~\ref{ex-set} we describe
the experimental setup and the samples. The experimental results are
reported in Secs.~\ref{ex-res} and \ref{ex-nl}. They are discussed
in Sec.~\ref{discussion}.

\section{Experimental procedure and results} \label{experimental}
\subsection{Experimental setup and sample} \label{ex-set}

As previously,~\cite{Drichko2015} we use  the so-called hybrid
acoustic method discussed in detail in
Ref.~\onlinecite{Drichko2000}, see Fig. 1 (left) in that paper. A
sample is pressed by springs to a surface of a LiNbO$_3$
piezoelectric crystal where two inter-digital transducers (IDTs) are
formed. One of the IDTs is excited by AC pulses. As a result, a SAW
is generated, which propagates along the surface of the
piezoelectric crystal. The piezoelectric field penetrates into the sample and the in-plane longitudinal component of the field interacts with the charge carriers. This interaction causes
SAW attenuation and deviation of its velocity.

 We study multi-layered $n$-GaAlAs/GaAs/GaAlAs structures
 with a wide (65 nm) GaAs quantum well (QW), the same as
 in Ref.~\onlinecite{Drichko2015} (see right panel of Fig.~1
 in that paper). The QW is $\delta$-doped from both sides
 and is located at the depth d = 845 nm from the surface.
 The electron density is $n=5\times 10^{10}$cm$^{-2}$  and
 the mobility is $\mu_{0.3\text{K}} = 8\times 10^6$ cm$^2$/V$\times$s. Studies show that at the given electron density only the lowest band of transverse quantization should be occupied.~\cite{Manoharan1996}

\subsection{Results: Linear response} \label{ex-res}

\begin{figure}
\centerline{
\includegraphics [width=\columnwidth]{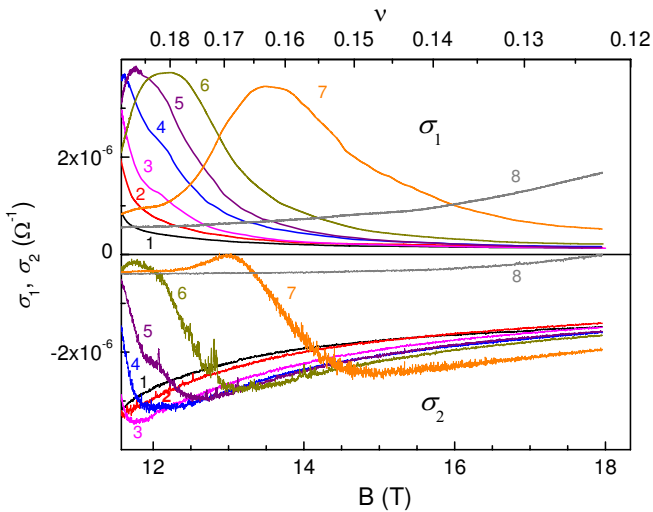}
}
\caption{(Color online) Magnetic field dependences of $\sigma_1$ and $\sigma_2$ for different temperatures, $T$, mK: 1 - 40, 2 - 74, 3 - 87, 4 - 100, 5 - 112, 6 - 122, 7 - 160, 8 - 240; $f=28.5$~MHz.
%
}
\label{fig1}
\end{figure}

Shown in Fig.~\ref{fig1} are  the magnetic  field dependences of
$\sigma_1$ and $\sigma_2$ for $f=28.5$~MHz extracted from
simultaneous measurements of the SAW attenuation, $\Gamma$, and
the relative variation of its velocity, $\Delta v/v$.
\begin{figure}
\centering
\includegraphics[width=\columnwidth]{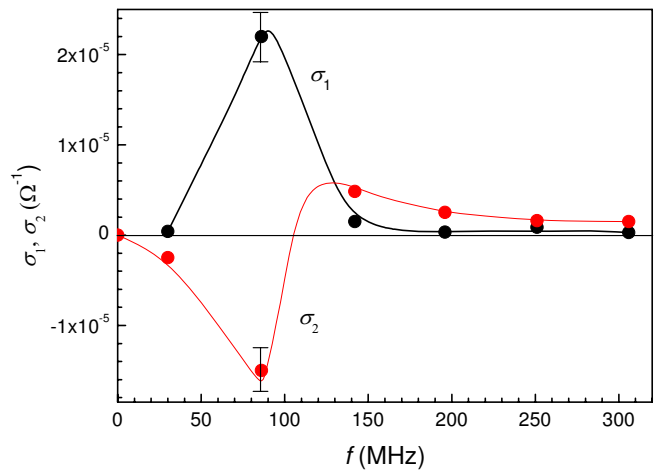}
\caption{Frequency dependences of $\sigma_1$ and $\sigma_2$ at $T=40$~mK and  $\nu=0.18$ ($B=12.2$~T).
Lines are guides for an eye.
\label{fig2}}
\end{figure}
These  data were used for calculating the components  of the complex
conductance, $\sigma ^{AC} (\omega) \equiv \sigma_1 (\omega)
-i\sigma_2 (\omega)$, according to the procedure outlined in
Ref.~\onlinecite{Drichko2000}. Namely, the complex AC conductance
$\sigma ^{AC} (\omega)$ was calculated using Eqs.~(1)-(7)
from~Ref.~\onlinecite{Drichko2000} where we substituted $\varepsilon_1=50$,
$\varepsilon_0=1$ and $\varepsilon_s=12$ for the dielectric
constants of the LiNbO$_3$ crystal, of the vacuum and of the sample,
respectively. The finite vacuum clearance $a = 5\times 10^{-5}$~cm
between the sample surface and the LiNbO$_3$ surface was determined
from the saturation value of the SAW velocity in strong magnetic fields at
$T=380$~K; $d=845$~nm is  the finite distance between the sample
surface and the 2DES layer. The SAW velocity is $v_0= 3\times
10^5$~cm/s.

The frequency dependences of the  components $\sigma_i$ shown in
Fig.~\ref{fig2} is a characteristic of the Wigner crystal pinned by disorder with pinning frequency $\sim
86$~MHz in this case.

\begin{figure}
\centerline{
\includegraphics [width=\columnwidth]{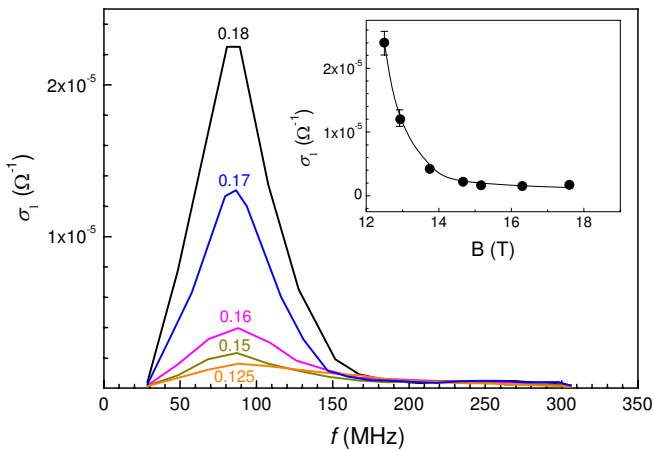}
}
\caption{(Color online) Frequency  dependences of $\sigma_1$
for different filling factors (shown near the curves).
Inset: Magnetic field dependence of
$\sigma_1(86~\text{MHz})$. $T$=40 mK.} \label{fig3}
\end{figure}

Shown in Fig.~\ref{fig3} are the frequency dependences of  $\sigma_1$
for different filling factors. The curves have maxima at $f \approx
86$~MHz, their amplitudes are decreasing when the magnetic field increases, see inset.

\begin{figure}[t]
\centerline{
\includegraphics [width=.9\columnwidth]{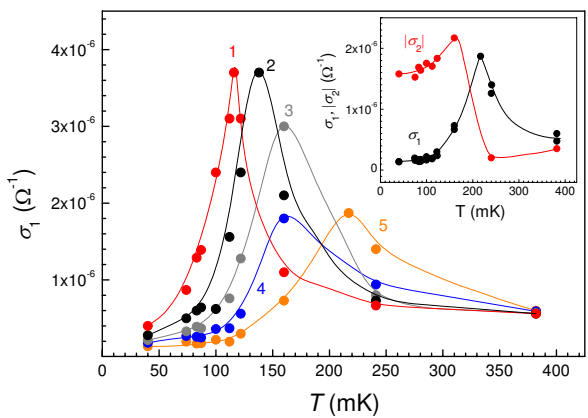}
}
\caption{(Color online) Temperature  dependences of $\sigma_1$
for different filling factors,
$\nu$: 1 - 0.18, 2 - 0.17, 3 - 0.16, 4 - 0.145, 5 - 0.125.
$f= 28.5$~MHz.
Inset: Temperature dependences of $\sigma_1$ and $\mid \sigma_2 \mid$
for $\nu$= 0.13,
$f= 28.5$~MHz. Lines are guides for an eye.
} \label{fig4}
\end{figure}

The same $\sigma_1$ data are presented in Fig.~\ref{fig4} as the temperature dependences at various
filling factors.
 The each curve has a maximum, which decreases  and shifts   towards
  higher temperatures with decrease of the filling factor.
  Such a behavior is also observed at other frequencies.

On the left of the maxima, the temperature  dependences of
$\sigma_1$ are clearly dielectric; in these regions $|\sigma_2| >
\sigma_1$. This fact, as well as the frequency dependences of
$\sigma_i$ in the magnetic field interval between 12 and 18~T
 can be attributed to a pinned mode of WC.
On the right of the maxima, $|\sigma_2|$ rapidly decreases with
temperature increase. $\sigma_1$ also decreases with temperature,
but much slower then $|\sigma_2|$, and at high temperatures the condition
 $|\sigma_2| < \sigma_1$ is valid.
Thus, it is natural to ascribe the maximum  - the temperature at which
the conduction mechanism rapidly changes - to the WC melting point,
$T_m$,  for a given filling factor.

\begin{figure}[t]
\centerline{
\includegraphics [width=.9\columnwidth]{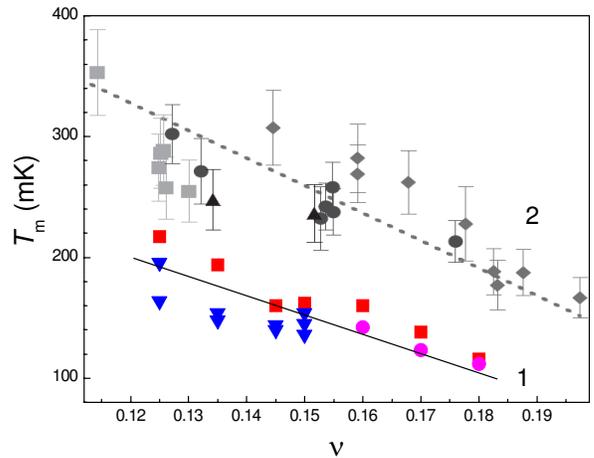}
}
\caption{(Color online)  1 - Dependence of the ``melting
temperature'', $T_m$, on the filling factor $\nu$ for
different frequencies, $f$:
$\blacksquare$ - 28.5~MHz, $\blacktriangledown$ - 86~MHz,
{\large $\bullet$}  - 142~MHz.  2 - Melting temperature, $T_m$,
from Ref.~\onlinecite{Chen2006} determined as the value where
resonances disappear.\label{fig5} }
\end{figure}
So-obtained dependences $T_m(\nu)$ for different frequencies are
shown in Fig.~\ref{fig5} as dataset 1. The dataset 2 presented in the same figure is taken from
Ref.~\onlinecite{Chen2006} where the temperature dependence
of the amplitude of the pinning resonance was studied as a function of the filling
factor in a GaAs/AlGaAs heterojunction with carrier density tuned by backgate in the range of $n = (1.2 \div 8.1) \times 10^{10}$ cm$^{-2}$. In Ref.~\onlinecite{Chen2006}, $T_m (\nu)$ was defined as the
temperature at which the pinning resonance disappears at a given
filling factor $\nu$. Note that the dependences $T_m(\nu)$ obtained in this research and in Ref.~\onlinecite{Chen2006}, i.e.,
by different procedures, are similar. However, the pinning resonance
disappears at higher temperature than the temperature where the
conduction mechanism rapidly changes. One can speculate that the
melting temperatures determined by different procedures correspond
to boundaries of the transition from the Wigner glass to the
electron liquid.

\subsection{Results: Nonlinear response} \label{ex-nl}

\begin{figure}
\centerline{
\includegraphics [width=.9\columnwidth]{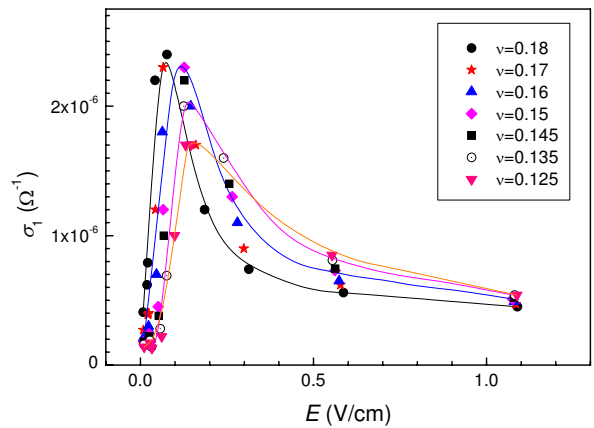}
}
\caption{(Color online)  Dependences of $\sigma_1$ on
the SAW electric field amplitude, $E$, for different
filling factors $\nu$.
$f=28.5$~MHz, $T=40$~mK.\label{fig6} }
\end{figure}

Shown in Fig.~\ref{fig6} are dependences  of $\sigma_1$ on the
amplitude of the electric field, $E$, produced by the SAW for
several filling factors $0.125 \le \nu \le 0.18$.
The electric field was determined  according to Eq.~(2) from
Ref.~\onlinecite{Drichko2000a}, see also
Ref.~\onlinecite{Drichko1997a}. The electric field dependences of $\sigma_1$ are similar to
the temperature dependences shown in Fig.~\ref{fig4}. Therefore,
increase in the SAW amplitude acts as an increase of the
temperature.

\begin{figure}[b]
\centerline{
\includegraphics [width=.9\columnwidth]{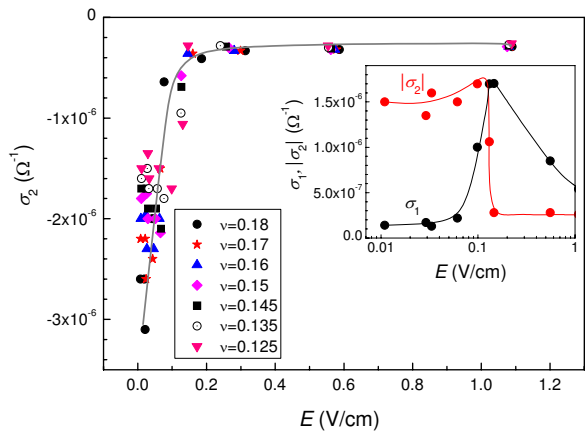}
}
\caption{(Color online)  Dependences of $\sigma_2 (E)$
for different filling factors $\nu$.
Inset: Dependences of $\sigma_1$ and $|\sigma_2|$ on the SAW electric field for $\nu = 0.125$.
$f=28.5$~MHz, $T=40$~mK. Lines are guides for an eye.
 \label{fig7} }
\end{figure}

The electric field dependences of $|\sigma_2|$ for  different $\nu$ are shown
in Fig.~\ref{fig7}.
Notice that the dependences $\sigma_2 (E)$ for different filling factors collapse on the same curve. For convenience, both components $\sigma_1$ and $|\sigma_2|$ at frequency 28.5~MHz and $\nu =0.125$ are presented in the same graph, see the inset. On the left of the maximum of $\sigma_1(E)$, $|\sigma_2| > \sigma_1$. This behavior is compatible with the predictions~\cite{Fogler2000} for a Wigner crystal. On the right of the maximum, $|\sigma_2|$ rapidly drops and becomes much less than $\sigma_1$. It indicates a change of the AC conduction mechanism.  Assuming that an intense SAW increases the temperature of the electron system we ascribe this behavior to melting of the Wigner crystal. The behaviors of $\sigma_{1,2}$ are similar for different frequencies with the exception of the frequency $f=142$~MHz at which $\sigma_2 >0$ at all used intensities.

 \section{Discussion} \label{discussion}

 The behavior of $\sigma (\omega)$ shown in Fig.~\ref{fig2} is typical for a pinned mode of a Wigner crystal,~\cite{Fertig1999,Yi2000,Fogler2000,Chitra2001,Fogler2004}  see also Refs.~\onlinecite{Shayegan1997,Shayegan2005} for a review. The crystal manifests itself in observed  resonances in $\sigma_1 (\omega)$,~\cite{Ye2002} which has been interpreted as a signature  of a solid and explained as due to the pinning mode (the disorder gapped lower branch of the magnetophonon)~\cite{Fukuyama1978,Normand1992,Fertig1999,Fogler2000,Chitra2001} of WC crystalline domains oscillating collectively within the disorder potential.
The WC states compete with the fractional quantum Hall effect (FQHE) states. Based on several experiments and calculations  it is concluded that at $\nu=1/5$ the FQHE dominates while at $\nu$ slightly less or slightly higher than 1/5 the WC state wins, see, e.g., Fig.~9 from Ref.~\onlinecite{Shayegan2005}.

The dynamic response of a weakly pinned Wigner crystal at  not too small frequencies is dominated by the collective excitations~\cite{Fertig1999,Fogler2000,Fogler2004} where an inhomogeneously broadened absorption line (the so-called pinning
mode) appears.~\cite{Fukuyama1977,Fukuyama1978} It corresponds to collective vibrations of correlated segments of the Wigner crystal around
their equilibrium positions formed by the random pinning potential. The mode is centered at some disorder- and magnetic-field-dependent frequency, $\omega_p$ (so-called pinning frequency), with a width being determined by a complicated interplay between different collective excitations in the Wigner crystal. There are modes of two types: transverse (magnetophonons) and longitudinal (magnetoplasmons). The latter include fluctuations in electron density. An important point is that pinning modifies both modes, and the final result depends on the strength and the correlation length, $\xi$, of the random potential. Depending in the strength and the correlation length of the random potential, the frequency, $\omega_p$ may either increase, or decrease when the magnetic field rises.

The ratio $\omega_p/\omega_c$, where $\omega_c$ is the cyclotron frequency, can be arbitrary. Depending on the interplay between the ratio $\omega_p/\omega_c$ and the ratio $\eta \equiv
\sqrt{\lambda/\beta}$ between the shear ($\beta$) and bulk ($\lambda$) elastic moduli of the Wigner crystal, one can specify two regimes where the behaviors of $\sigma^{\text{AC}}$ are different:
\begin{equation} \label{regimes}
(a) \ 1 \ll \omega_c/\omega_{p0} \ll
\eta, \quad (b) \ 1 \ll \eta
  \ll \omega_c/\omega_{p0}     \, .
\end{equation}
Here $\omega_{p0}$ is the pinning frequency at $B=0$. As a result, the
variety of different behaviors is very rich.
 Assuming $\xi \gg
l_B=(\hbar c/eB)^{1/2}$ one can cast the expression for
$\sigma_{xx}(\omega)$ from Ref.~\onlinecite{Fogler2000} into the form
\begin{equation}
  \sigma (\omega)=-i\frac{e^2n\omega}{m^*\omega_{p0}^2}\frac{1-iu(\omega)}
{[1-iu(\omega)]^2   -(\omega
  \omega_c/\omega_{p0}^2)^2}\, , \label{sigma1}
\end{equation}
where the function $u(\omega)$ is different for regimes (a) and (b).

Let us consider the regime (b) since only this regime seems to
be compatible with our experimental results. Then
\begin{equation}
  \label{eq:fb}
  u(\omega) \sim \left\{ \begin{array}{lll}
(\omega/\Omega)^{2s}, & \omega \ll \Omega\, , &\quad  (b1)\\
\text{const}, & \Omega \ll \omega \ll \omega_c \, .& \quad (b2) \end{array}
\right.
\end{equation}
Here $\Omega \sim \omega_{p0}^2\eta/\omega_c$, while $s$ is some
critical exponent. According to Ref.~\onlinecite{Fogler2000}, $s=3/2$.

\begin{figure}
\centering
\includegraphics[width=0.8\columnwidth]{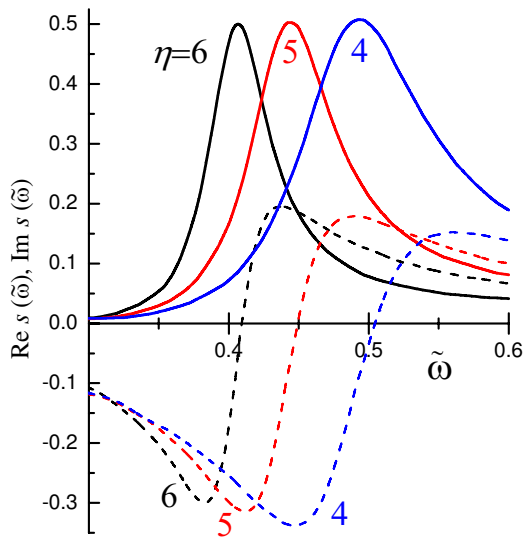}
\caption{(Color online) Graphs of $\Re s$ (solid lines) and
$\Im s$ (dash  lines) for $\eta = 4$, 5, and 6. \label{fig8}}
\end{figure}

Assuming the regime (b1) we can cast Eq.~(\ref{sigma1}) in the form
 $\sigma(\omega)\equiv \sigma_0  s(\omega/\Omega)$
where
\begin{equation}
\label{eq:fb4}
\sigma_0 \equiv \frac{e^2n  \eta^2}{2m^*\omega_c}\,,  \quad
s(\tilde{\omega})=-2\frac{i \tilde{\omega}(1-i \tilde{\omega}^3)}
{\eta[(1-i \tilde{\omega}^3)^2- (\eta \tilde{\omega})^2]}\, ,
\end{equation}
with $\tilde{\omega}=\omega / \Omega$. This function is normalized
in order to have its maximum $\eta$-independent. Graphs of real and
imaginary parts of $s(\omega/\Omega)$ for $\eta=4$, 5 and 6 are
shown in Fig.~\ref{fig8}.

Equation~(\ref{eq:fb4}) predicts  decrease of the maximum magnitude
of $\sigma_1 (\omega)$ with increase of magnetic field. This
prediction is compatible with our experiment, see the inset in Fig.~\ref{fig3}.  However, the predicted behavior of maximum frequency as
$\omega_p \propto \omega_c^{-1}$  is not observed -- the resonant
frequency is almost independent of magnetic field, as seen in
Fig.~\ref{fig3}. One needs to note, however, that the specificity of
our experimental technique does not allow to trace the impact of
small change in frequency on the dependence $\sigma$ on $\omega$.

Unfortunately, the experimental data  shown in Fig.~\ref{fig2} do
not provide an accurate structure of the maximum, and therefore do
not allow fitting the model with high accuracy. Assuming $\eta =5$
that gives approximately correct shape of the curves in
Fig.~\ref{fig2}      and taking into account that
the maximum of $\sigma_1(\omega)$  occurring at $\omega_{\max}/2\pi
\approx 86$ MHz   corresponds to $\omega /\Omega =0.44$, we conclude
that $\Omega =\omega_{\max}/0.44 \approx 1.2\times 10^9$~s$^{-1}$.
The quantity $\omega_p \equiv 0.44\Omega_{\max}=0.44\omega_{p0}^2
\eta/\omega_c$ plays the role of the pinning frequency in the magnetic
field.~\cite{Fogler2000}

The frequency $\omega_{p0}$ can then be  determined as
\begin{eqnarray*}
\omega_{p0}=\sqrt{\omega_c \Omega/\eta}.
\end{eqnarray*}
Substituting $\eta=5$, $\Omega =  1.2\times 10^9$~s$^{-1}$,
$\omega_c= 3.2\times 10^{13}$~s$^{-1}$ ($B=12.2$~T, $\nu =0.18$) we
obtain
\begin{eqnarray*}
\omega_{p0} =8.7\times 10^{10}~\textrm{s}^{-1}.
\end{eqnarray*}
Therefore, the regime (b) of
Eq.~(\ref{regimes}) is the case, as we expected.

Estimating the Larkin
 length, i.e., the WC domain correlation length, as \begin{eqnarray*}
L =2 \pi c_t/\omega_{p0},
\end{eqnarray*}
where $c_t  = (\beta/nm^*)^{1/2} \approx 4\times 10^6$~cm/s is the
velocity of the WC transverse mode for our electron density $n$ we
obtain $L \approx 3\times 10^{-4}$~cm that is much larger than both
the distance between the electrons $a=4.8 \times 10^{-6}$~cm and the
magnetic length $l_B=7.3 \times 10^{-7}$~cm,
\begin{eqnarray*} L \gg a \gg l_B.
\end{eqnarray*}
These inequalities justify
using the theory~\cite{Fogler2000} for our estimates.

In conclusion, we have measured the absorption and the velocity of SAWs in
high-mobility samples $n$-GaAs/AlGaAs in magnetic fields 12$\div$18~T
(i.e., at filling factors $\nu = 0.18\div0.125$). From the measurement
results the complex AC conductance, $\sigma ^{AC} (\omega) \equiv
\sigma_1 (\omega) -i \sigma_2 (\omega)$ was found, and its
dependences on frequency, temperature and the amplitude of the
SAW-induced electric field were discussed. We conclude that in the
studied interval of the magnetic field and $T < 200$~mK the electronic
system forms a pinned Wigner crystal, the so-called Wigner glass. The
estimate of the correlation (Larkin) length of the Wigner glass is
$\approx 3$~$\mu$m at $B=12.2$~T.

We have also  defined an effective melting temperature,  $T_m$, as
the temperature corresponding to the maximum in the temperature
dependence of $\sigma_1$, or rapid decrease of $|\sigma_2|$. These
behaviors indicate a rapid change in the conductance mechanism -- from
the dielectric behavior at $T<T_m$
to the metallic one at $T>T_m$.

\begin{acknowledgments}
I.L.D. is grateful for support from RFBR via grant 14-02-00232. The
authors would like to thank E. Palm, T. Murphy, J.-H. Park, and G.
Jones for technical assistance. NHMFL is supported by NSF Cooperative Agreement
DMR-1157490 and the State of Florida. The work at Princeton
University was funded by the Gordon and Betty Moore Foundation
through the EPiQS initiative Grant GBMF4420, and by the National
Science Foundation MRSEC Grant DMR-1420541.
\end{acknowledgments}

\end{document}